\documentclass[sigconf, nonacm]{acmart}

\newcommand\vldbdoi{XX.XX/XXX.XX}
\newcommand\vldbpages{XXX-XXX}
\newcommand\vldbvolume{14}
\newcommand\vldbissue{1}
\newcommand\vldbyear{2020}
\newcommand\vldbauthors{\authors}
\newcommand\vldbtitle{\shorttitle} 
\newcommand\vldbavailabilityurl{URL_TO_YOUR_ARTIFACTS}
\newcommand\vldbpagestyle{plain} 

\begin{document}
\title{A Demonstration of Benchmarking Time Series Management Systems in the Cloud}

\author{Prabhav Arora}
\affiliation{%
  \institution{New York University Abu Dhabi}
  \streetaddress{P.O. Box 1212}
  \city{Abu Dhabi}
  \state{UAE}
}

\begin{abstract}
Time Series Management Systems (TSMS) are Database Management Systems that have been configured with the primary objective of processing and storing time series data. With the IoT expanding at exponential rates and there becoming increasingly more time series data to process and analyze, several TSMS have been proposed and are used in practice.  Each system has its own architecture and storage mechanisms and factors such as the dimensionality of the dataset or the nature of the operators a user wishes to execute can cause differences in system performance. This makes it highly challenging for practitioners to determine the most optimal TSMS for their use case. To remedy this several TSMS benchmarks have been proposed, yet these benchmarks focus primary on simple and supported operators, largely disregarding the advanced analytical operators (ie. Normalization, Clustering, etc) that constitute a large part of the use cases in practice. In this demo, we introduce a new benchmark that enables users to evaluate the performance of four prominent TSMS (TimescaleDB, MonetDB, ExtremeDB, Kairos-H2) in their handling of over 13 advanced analytical operators. In a simple and interactive manner, users can specify the TSMS(s) to compare, the advanced analytical operator(s) to execute, and the dataset(s) to utilize for the comparison. Users can choose from over eight real-world datasets with varying dimensions or upload their own dataset. The tool then provides a report and recommendation of the most optimal TSMS for the parameters chosen.
\end{abstract}

\maketitle

\pagestyle{\vldbpagestyle}
\begingroup\small\noindent\raggedright\textbf{PVLDB Reference Format:}\\
\vldbauthors. \vldbtitle. PVLDB, \vldbvolume(\vldbissue): \vldbpages, \vldbyear.\\
\href{https://doi.org/\vldbdoi}{doi:\vldbdoi}
\endgroup
\begingroup
\renewcommand\thefootnote{}\footnote{\noindent
This work is licensed under the Creative Commons BY-NC-ND 4.0 International License. Visit \url{https://creativecommons.org/licenses/by-nc-nd/4.0/} to view a copy of this license. For any use beyond those covered by this license, obtain permission by emailing \href{mailto:info@vldb.org}{info@vldb.org}. Copyright is held by the owner/author(s). Publication rights licensed to the VLDB Endowment. \\
\raggedright Proceedings of the VLDB Endowment, Vol. \vldbvolume, No. \vldbissue\ %
ISSN 2150-8097. \\
\href{https://doi.org/\vldbdoi}{doi:\vldbdoi} \\
}\addtocounter{footnote}{-1}\endgroup

\ifdefempty{\vldbavailabilityurl}{}{
\vspace{.3cm}
\begingroup\small\noindent\raggedright\textbf{PVLDB Artifact Availability:}\\
The source code, data, and/or other artifacts have been made available at \url{https://github.com/prabhav2302/ABench-IoT-Demo}.
\endgroup
}

\section{Introduction}
Time series data is essentially a collection of measurements of a certain quantity obtained at different moments in time. With the rapid expansion of the Internet of Things (IoT), such data has become increasingly more available and prevalent in many domains of industry and research. A Time Series Management System (TSMS) is a Database Management System that has been specifically formulated and optimized for processing, storing, and interacting with time series data. Each TSMS has its own architecture and storage mechanism which determines how data is represented and stored internally within the system. TSMS architects make several design decisions that in turn impact the overall performance of their system. 

One primary decision is whether to have a row-oriented (storing entries of a row in contiguous memory locations) or column-oriented (storing entries of a column in contiguous memory locations) system. Another decision designers must make is how to structure related time series. Some TSMS store each time series independently whereas other group subsets or all related time series together in storage. Though these decisions typically create differences between the architecture of the systems, one common property of most TSMS is that they partition data into chunks along the time axis. This representation results in faster computation times as the data can be pruned more efficiently since the work is parallelized across the time intervals. 

In addition to the system design, each TSMS must also provide support for various operators that allow users to interact with the data. The complexity of these operators is highly dependent on the system architecture and design as they dictate how each operator must be run. Certain architectures, such as row-store systems, favor certain operations, such as inserting and reading, whereas other architectures, such as column-store systems, favor other operations, such as aggregation calculations.  Aside from the operators a user wishes to utilize, the dataset a user wishes to interact with also impacts the performance of the TSMS. Once again certain TSMS architectures are more tailored towards datasets of certain properties (granularity, anomalies, cardinality). 

These permutations and combinations of desired user operators and datasets all result in different TSMS performance and thereby mitigate the existence of a blanket, silver bullet, “best” TSMS. How then does a practitioner navigate the performance of these TSMS and choose the most ideal TSMS for their use case?

Multiple benchmarks have been devised that contrast the runtime, memory usage, architecture, or quality of various TSMS. These evaluations typically begin with inserting a certain time series dataset into the respective TSMS and then then querying the system using a fixed set of operations. These benchmarks, however, largely concern themselves with applying simple and supported operators, and fail to take into consideration the advanced analytical operators which constitute a large portion of the use cases in time series applications. The performance of TSMS executing operators that do Reading and Inserting have been widely studied whereas the performance of executing operators that do Normalization, Clustering, and Mapping has been scarcely documented. 

AIoT-Bench aims to be an interactive and easy-to-use tool that enables users to examine the performance of several TSMS whilst executing various analytical operators on a diverse set of datasets. Users can evaluate the performance of eight prominent TSMS (InfluxDB, TimescaleDB, MonetDB, ExtremeDB, Graphite, Kairos-Cassandra, Kairos-H2, Druid) in their handling of over 9 advanced analytical operators (Normalize, Map, Decompose, Recover, Repair, Cluster, Classify, Pattern, Similarity) and 4 simple analytical operators (Records, SumRecords, MovingAverage, Distance). Users have access to 6 real-world datasets, each with its own unique and distinct properties, that have been specifically selected to emulate and encompass the properties of data observed in practice. Users may also upload their own dataset to compare TSMS performance for their specific use case. After a user specifies exactly which TSMS to compare, which analytical operator to execute, and which dataset to utilize, AIoT-Bench provides a report and recommendation of the most optimal TSMS for the parameters chosen.

\section{Analytical Queries}
Our benchmark supports over 13 analytical operators, which can be further subdivided into 5 overarching types of operators. Below is an explanation of each operator and its potential use cases. 

\subsection{Advanced Transformation Operators}
Transformation operators, are those operators which when given data in one state perform some transformations and return the data in another, more useful or insightful state. Our benchmark supports Centroid Decomposition, SaxRepresentation, ZNormalization, and ZNormalization-Operators. 
\\{\bf Centroid Decomposition} is an iterative algorithm that decomposes an input matrix X into a product of two or more matrices. CD can be used to extract the principal features of a time series.
\\{\bf SaxRepresentation} first normalizes the data using ZNormalization, then reduces its dimensionality using Piece-wise Aggregate Approximation. Then, values are assigned to a set of pre-defined buckets and mapped to a lower-case letter based on which bucket they belong to. 
\\{\bf ZNormalization} first calculates the standard deviation and mean of the data. Then for each point the Z-Score is calculated. Z-Score for $x$ = ($x$ - Mean) / Standard Deviation. Z-score normalization is used in most time series analytical tasks as a preprocessing step. Our benchmark tests Z-normalization using both UDF and the in-built operators of each database.

\subsection{Machine Learning Operators}
Machine Learning operators, are operators which perform a ML algorithm on a dataset. Our benchmark supports KMeans and KNN.
\\{\bf KMeans} is an unsupervised learning algorithm which iteratively clusters the dataset into $n$ groups and retrieves $n$ centroids.
\\{\bf KNN} is a supervised learning algorithm which classifies each point in the dataset to a category from a finite set based on its euclidean distance to the nearest neighbour from each category. 

\subsection{Erroneous Data Handling Operators}
Erroneous Data Handling Operators are operators which deal with missing or anomalous data. Our benchmark supports Screen, Recovery, and HotSax operators.
\\{\bf Screen} is a technique that detects anomalies and repairs them. 
\\{\bf Recovery} is implemented using RecovDB\cite{8731357}. The input data has missing values and output data estimates the missing values and adds them to the initial set. 
\\{\bf HotSax}\cite{1565683} processes each column at a time and looks for anomalies by first normalizing the data and then applying euclidean distance and a SAX representation. It returns a number $x$ of anomalies for time series  within a given time interval

\subsection{Simple Operators}
In order to provide a comprehensive and an all-inclusive benchmark, Our benchmark also supports basic operators such as Select and Sum.
\\{\bf Select} the datapoints (observations) whose timestamps lie in a given window. Provides a way for the clients to retrieve the data within the desired time-range. 
\\{\bf Sum} computes an aggregate over the entire time series using the SUM operator. 

\subsection{Similarity Detection Operators}
Similarity Detection operators are operators which identify or quantify the degree of similarity between time series. Our benchmark supports Distance and DSTree. 
\\{\bf Distance} computes the Eucledian distance between the time corresponding points in time series $ts$1 and $ts$2. Is the building block for more complex operators like KNN, and KMeans.
\\{\bf DSTree}\cite{4053129} retrieves for each time series in set $ts$2, the closest time series in set $ts$1. (Closeness defined by Euclidean distance between two time series). The algorithm is split into two parts: the indexing (using DS-tree) and the actual search based on the index.

\begin{figure}[h!]
  \includegraphics[scale=0.27]{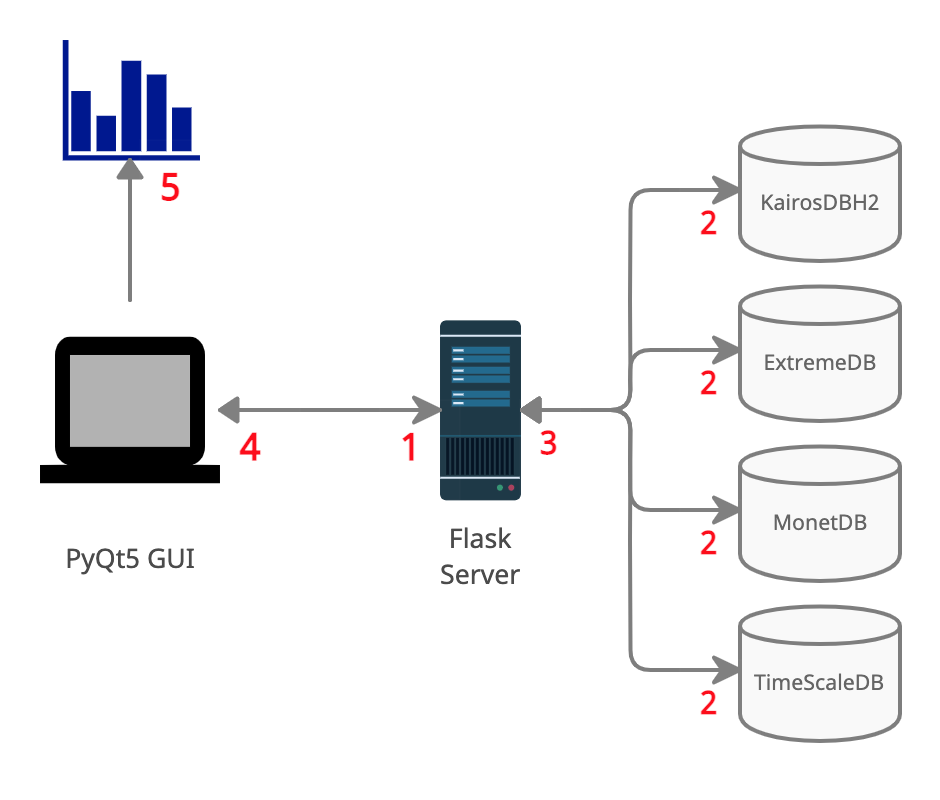}
  \caption{Architecture Diagram}
  \label{fig:architecture}
\end{figure}

\section{Benchmark GUI}
Our benchmark has a graphical user interface implemented with the PyQt5 framework for the front end, and Flask server framework for the back end. The architecture of Our benchmark can be visualized in Figure 1. In step 1, a client interacts with the GUI and selects the parameters and clicks run, which prompts the GUI to send a HTTP GET request to the Flask server. The remote machine which the Flask server is hosted on has all the prerequisite databases installed and running. In step 2 the Flask server handles the request and executes the relevant scripts on the Databases as specified by the client's parameters. In step 3, each TSMS sends its runtime results back to the Flask server. The Flask server then sends the GUI a CSV file containing the results in step 4. Finally, the GUI parses the CSV file and plots a graph with the results.

\begin{figure}[h!]
  \includegraphics[scale=0.35]{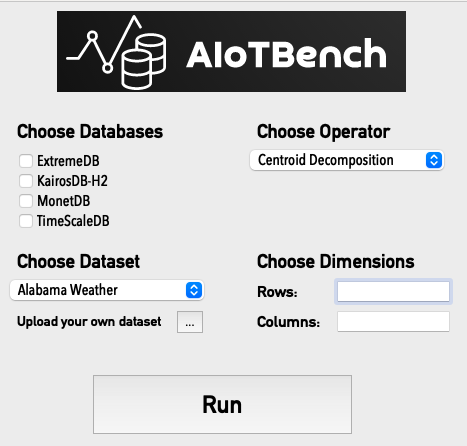}
  \caption{Our benchmark PyQt5-based GUI}
\end{figure}

\subsection{Supported Operators}
Our benchmark implements all the 13 operators discussed above, enabling users to test the performance of the various databases on a wide spectrum of use cases. Each operator script contains a UDF to execute the specified operator. For Our benchmark all UDFs were coded using Python to ensure consistency and fairness in experiments. 

\subsection{Datasets}
In addition to the user being able to upload custom datasets, Our benchmark comes preloaded with 8 datasets. Each preloaded dataset exhibits different characteristics of data (dimensionality, anomalies, regularity, etc.) from the other. The decision to include datasets with distinct differences has been made consciously in order to account for the various use cases of data one can expect to encounter in real life. 
\begin{center}
\begin{tabular}{||c c c c||} 
 \hline
 Name & Number of TS & Length of TS & Features \\ [0.5ex] 
 \hline\hline
 Alabama & 46 & 3.500.000 & anomalies, regular \\ 
 \hline
 Sports & 360 & 142.500 & regular, anomalies \\
 \hline
 MexData & 512 & 7.000 & irregular, anomalies \\
 \hline
 Hydraulic & 43.680 & 2.205 & regular,anomalies \\ [1ex] 
 \hline
\end{tabular}
\end{center}

 The Alabama weather dataset\cite{Refworks:123} has been included due its large magnitude and large length of time series (3.5m). The Sports dataset\cite{Dua:2019} was chosen as a middle-way dataset, with a medium number of time series (360) and a medium-large length (140k). Next, the MexData\cite{refworks:901}  is selected as an irregular dataset of a medium number of time series and small-medium length. Finally, the Hydraulic dataset\cite{8666564} was chosen due to its very large number of time series, small-medium length, and large total volume of data.  The other four preloaded datasets, also display varying data characteristics to those mentioned above. 

\begin{figure}[h!]
  \includegraphics[scale=0.35]{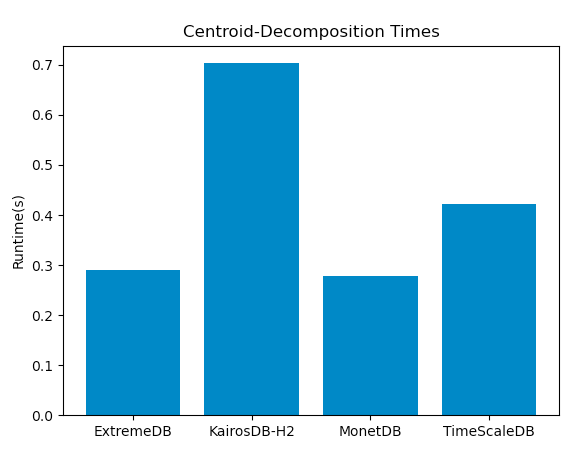}
  \caption{Results Plot}
  \label{fig:results}
\end{figure}

\section{Demonstration Scenarios}
{\bf Scenario 1: Choosing A Preloaded Dataset}. In this scenario a user must first select which databases he/she wishes to execute queries on. Then, the user must decide which operator they wish to use. Next, the user must choose from one of the 8 pre-loaded datasets available on the benchmark. Then, they must specify exactly how many rows and columns of the dataset, they want the benchmark to take into consideration. As soon as the user hits the "Run" button, a HTTP GET request is sent to the Flask server and after the queries are executed on the remote machine, the results are returned and plotted for the user. Figure 3 shows the results of Our benchmark for Centroid Decomposition on 1000 rows and 10 columns of the Alabama dataset. Subsequent to retrieving the results of a query, the user may tinker with the parameters and further execute more tests. 
\\{\bf Scenario 2: Uploading a Custom Dataset}. The workflow for this scenario is identical to Scenario 1, except in the part where the user has to select a Dataset. Instead of choosing one of the 8 preloaded Datasets, a user must instead choose "User Dataset" on the Dataset drop down menu. Then, the user should click the button with 3 dots and choose which dataset file from their machine they want to use. Subsequent to retrieving the results of a query, the user may tinker with the parameters and further execute more tests.

\begin{acks}
 This work was supported by the eXascale Infolab in Switzerland. We also thank Professor Djellel Difallah for his mentorship and oversight of this research project.
\end{acks}


\bibliographystyle{ACM-Reference-Format}
\bibliography{bib}

\end{document}